\documentstyle[11pt,newpasp,twoside,epsf]{article}
\def\edcomment#1{\iffalse\marginpar{\raggedright\sl#1\/}\else\relax\fi}
\reversemarginpar

\begin{document}
\title{SHEEP: the ASCA 5-10 keV survey}
 \author{I. Georgantopoulos}
\affil{Institute of Astronomy \& Astrophysics, National Observatory 
of Athens, Greece}
\author{K. Nandra}
\affil{Goddard Space Flight Center, NASA, Greenbelt, Maryland, USA}
\author{A. Ptak}
\affil{Physics Department, Carnegie Mellon University, Pittsburgh, USA}

\begin{abstract}
We present the first results of the hard (5-10 keV) 
 ASCA GIS survey SHEEP (Search for the High Energy Extragalactic Population).
 We have analysed 149 fields covering an area of 39 deg$^2$ 
 detecting 69 sources. Several of these
 appear to be associated with QSOs and Seyfert-1 galaxies 
 but with hard X-ray spectra, probably due to high 
 absorption ($N_H>10^{22} \rm cm^{-2}$) 
 Indeed, the hardness ratio analysis shows that 
 the spectra of the majority of our sources can 
 be represented with a ``scatterer'' model
 similar to obscured Seyfert galaxies locally. 
 According to this model, our sources present  
 high intrinsic absorption ($N_H\sim10^{23}$) but
 also significant amounts of soft X-ray emission 
 coming from scattered light. 
 \end{abstract}

\section{Introduction}
The X-ray background (XRB) is made up of emission 
by discrete sources. Previous soft (0.1-2 keV) ROSAT X-ray surveys 
 demonstrated that a large fraction of the XRB constitutes of 
 QSOs (Shanks et al. 1991, Hasinger et al. 1998). 
 Harder X-ray surveys in the 2-10 keV band with ASCA and BeppoSAX 
 suggested that obscured AGN also contribute 
 a large fraction of the XRB (Georgantopoulos et al. 1997, 
  Cagnoni et al. 1998, Ueda et al. 1998, Akiyama et al. 2000,
 Giommi et al. 2000).
  These obscured AGN are associated with either 
 Seyfert type galaxies locally, or with broad-line AGN (QSOs) 
 at high redshift (Georgantopoulos et al. 1999, Fiore et al. 1999).
 Recently, Chandra observations (eg Mushotzky et al. 2000,
  Giacconi et al. 2000) have almost resolved the XRB in the 2-10 keV band 
  in a small area of the sky and hence with very small number statistics.
 Surprisingly, the Chandra observations have increased the 
 mystery of the X-ray background even further. 
 Mushotzky et al. find two new populations: one is associated with 
 anonymous early-type galaxies while the 
 other one has no or extremely faint (B$>$25) optical counterparts.
 Both present very hard X-ray spectra harder than the spectrum of the 
 XRB.  

 In order to identify the nature of these enigmatic hard sources 
 one needs to find brighter examples for spectroscopic follow-up 
 in both optical and X-ray wavelenghts. Unfortunately, 
 the Chandra surveys contain sources with typically 
 only a few tens of photons and they do not cover 
  sufficient area to reveal rare, bright examples of the 
 above populations. We have therefore begun a large area (40 $\rm deg^{2}$)
 hard (5-10 keV) X-ray survey with the ASCA GIS to find such 
 examples in the nearby Universe. 
 Our ASCA survey SHEEP (Search for the High Energy Extragalactic Populations) 
 is very similar to the BeppoSAX HELLAS survey (Fiore et al. 1999). 
  We believe that both the SHEEP and the HELLAS surveys 
 are timely complements for the deep Chandra surveys 
 which will pin down the brightest examples of the long sought high 
 redshift AGN population and the new classes of Chandra objects.

\section{The SHEEP survey}
We have chosen to use the GIS instrument onboard ASCA 
 due to its large FOV (0.33 deg$^2$) and high sensitivity 
 in the 5-10 keV band. We are using fields from the 
TARTARUS AGN database (Turner et al. 2000). 
 We reject fields with a) exposure 
 less than 30 ksec (combined GIS2+GIS3) 
 b) Galactic latitude $|b|<20$ and c) target 
 brighter than 0.02 $ct~s^{-1}$  
 (as a bright target could contaminate the whole field
 due to the extended wings of the PSF ). 
 Our survey consists of 149 fields covering an area 
 of 39$\rm deg^2$. 
 We find candidate sources in the coadded GIS2 and GIS3 images
 by simply eyeballing the images (on average there are about 
 0 or 1 sources per image). We then use a circular 
 detection cell to determine the significance 
 of the detection. 
 We have detected 69 sources with a significance 
 above a Poisson probability threshold 
 of $3\times 10^{-5}$ or (equivalently 4.5$\sigma$)  
 in a ``detection'' cell of $\sim4$ arcmin$^2$. 
 Our faintest source  has a count  rate of 
 $\sim7\times10^{-4} cts~s^{-1}$
 corresponding to a flux of $9\sim10^{-14}$  (5-10 keV) 
 ($\Gamma=1.6$).

We have cross-correlated our catalogue with existing 
 X-ray and optical catalogues. 
 59 SHEEP sources are located within the 2-degree diameter FOV 
 of a pointed ROSAT PSPC observation. However, only 
 32 have ROSAT counterparts within 2 arcmin (2 $\sigma$ 
 GIS position error). 
 5 of them have a ROSAT all-sky survey  
 Bright Source Catalogue (RASSBSC) counterpart.
  By offseting the positions of the SHEEP sources, we 
 find that the number of ROSAT (WGACAT)  
 and ASCA source chance coincidences 
 is less than 3 (it is practically nil for the RASSBSC 
 sources).
  A large number of the 
 sources with ROSAT counterpart have already spectroscopic classification 
 in the literature. In table 1 we give the 
 optical -IDs of these sources. 
It is evident that 
 the dominant population is  type-1 AGN ie QSOs 
 and Seyfert-1 galaxies. Our highest redshift source 
 is the QSO RXJ125827+3528.0 at z=1.88.

\begin{table}
\begin{center}
\caption{The ROSAT subsample: optical IDs from the literature}
\begin{tabular}{cc} 
\hline 
QSOs/Sy1 & 10 \\
Sy2/LINER& 3 \\
Radio Gal.& 1 \\
Cluster & 1 \\
No-ID &  16  \\
\hline 
\end{tabular}
\end{center}
\end{table}

 In Fig. 1 we are plotting the number count distribution, logN-logS,
  in the 5-10 keV band for our sample. We have used a spectrum of $\Gamma=1.6$ 
 to convert count rates to flux.  
 We are comparing with the  logN-logS derived from the BeppoSAX 
 HELLAS survey (Comastri et al. 2000). We are also 
 comparing with the ASCA 2-10 keV logN-logS (Cagnoni al. 1998)
 converted in the 5-10 keV band again assuming a
 mean source spectrum of $\Gamma=1.6$. 
  We see that the Cagnoni et al. logN-logS, 
  $N(>S)\approx 10^{-21} S^{-1.67}$ (in the 5-10 keV band)
 provides a good fit to both the SHEEP and HELLAS data. 
 At the faintest flux probed by our survey we are 
 resolving about 15\% of the 5-10 keV XRB (using the HEAO-1 XRB 
 normalization of Marshall et al. 1980). 

\begin{figure}[t]
\plotfiddle{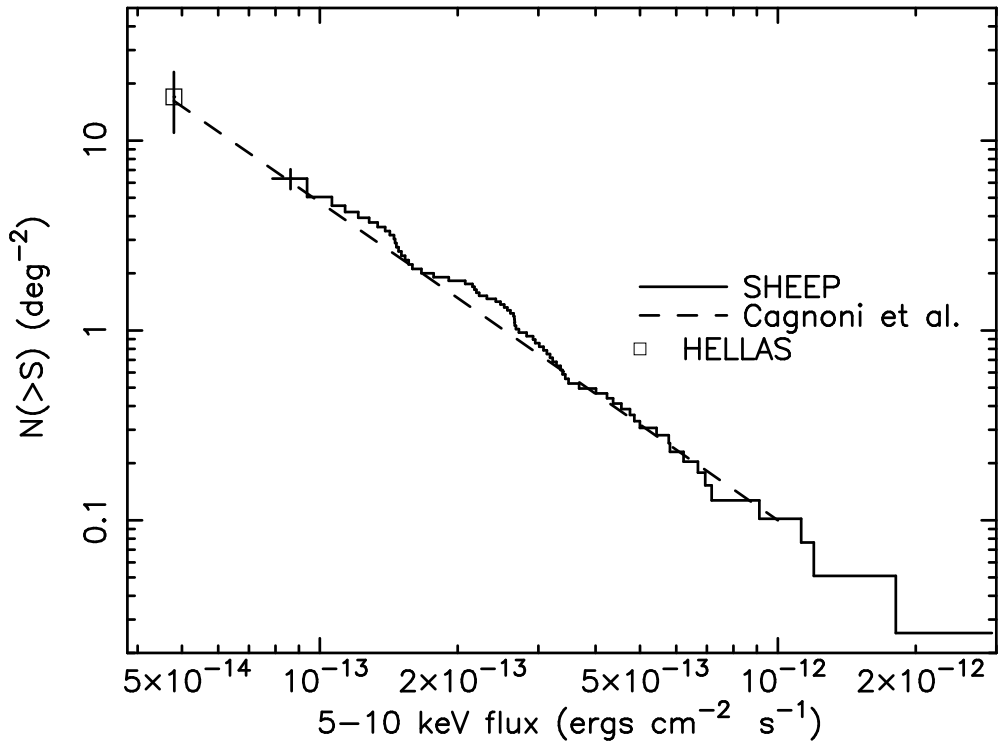}{10truecm}{270}{57}{57}{-210}{330}
\caption{ The 5-10 keV SHEEP logN-logS  (solid line) compared to the 
 BeppoSAX HELLAS survey (open square) and the ASCA 2-10 keV 
 logN-logS (Cagnoni et al. 1998) converted in the 
 5-10 keV band using $\Gamma=1.6$}
\end{figure}

\section{The spectral properties}

In Fig. 2 we plot the hardness ratio  versus the count rate 
 in the 5-10 keV band for our 69 sources. The hardnes ratio 
 is defined as h-m/h+m where h and m are the counts
 in the 2-5 and 5-10 keV band respectively
 (corrected for vignetting and light falling outside the 
 detection cell). 
 The solid line denotes a spectrum with $\Gamma=1.4$ 
 ie the spectrum of the XRB in the ASCA  band 
 (Gendreau et al. 1995).
We see that  large fraction of our 
 sources have spectra harder than that of the XRB.
 There is also a separation between the sources detected 
 by ROSAT (solid circles) and the sources with ASCA detections only  
 (open circles) with the former being in general
 softer than the latter. 
 Interestingly, some of the ROSAT sources with 
 hard X-ray specra ($\Gamma<1.4$ )
 are high redshift QSOs (eg CCRS 1429.7+4240)
 We note that the 2-10 keV spectrum does not get harder 
 with decreasing flux. 
 This is in contrast to previous X-ray spectral studies 
 of sources detected in the 2-10 keV band (Della Ceca et al. 
 1999, Giommi et al. 2000).

\begin{figure}[t]
\plotfiddle{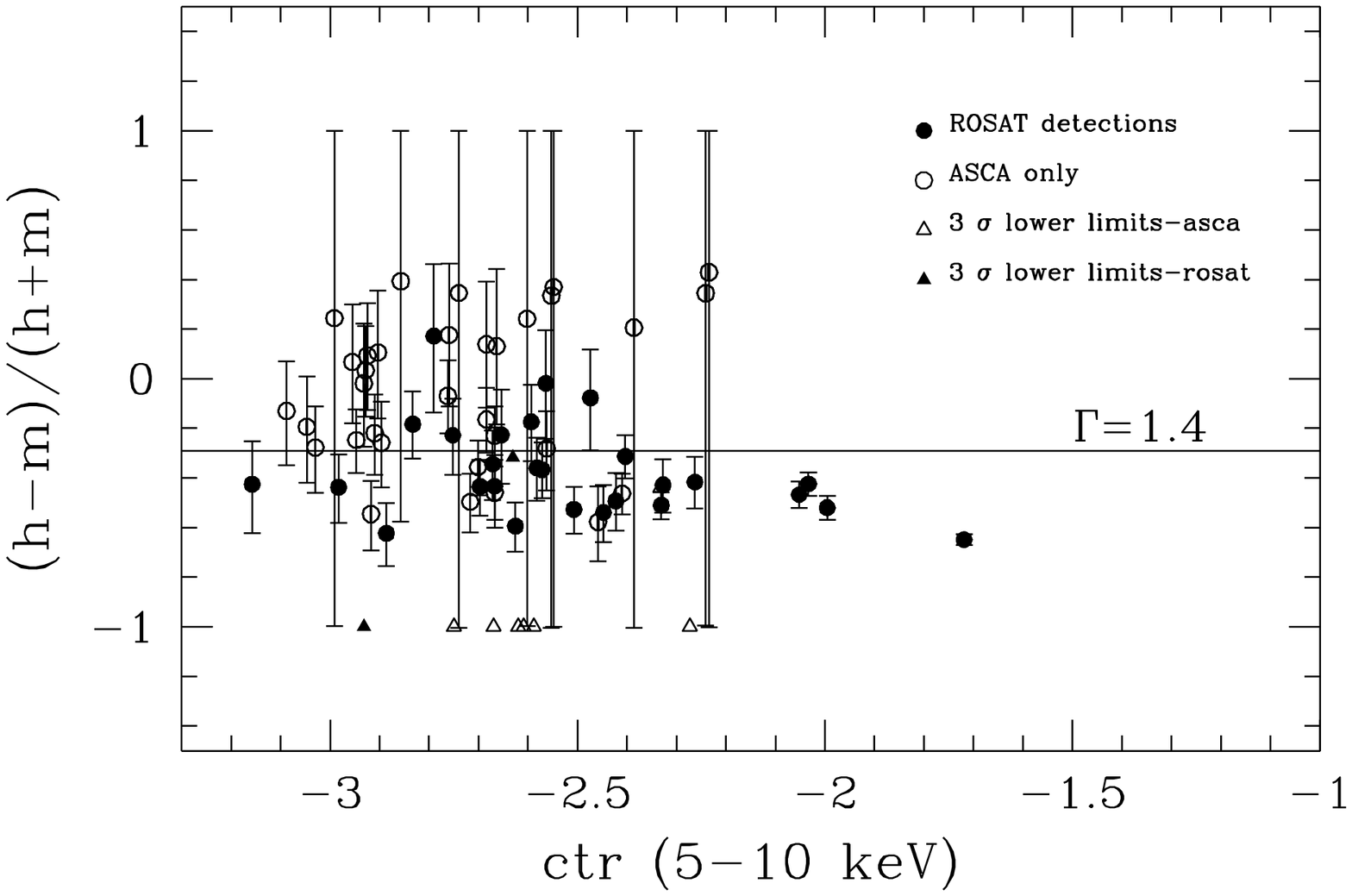}{10truecm}{0}{75}{75}{-250}{-120}
\caption{ The hardness ratio in the 2-10 keV band. 
 h and m denote the counts in the 2-5 and the 5-10 keV band respectively.
 Filled circles 
 correspond to sources with ROSAT detections while open circles 
 to those with ASCA detections only. Triangles denote 
 sources (3$\sigma$ upper limits) with no significant detection 
 in the 2-5 keV band.}
\end{figure}

\vfil

\begin{figure}[t]
\plotfiddle{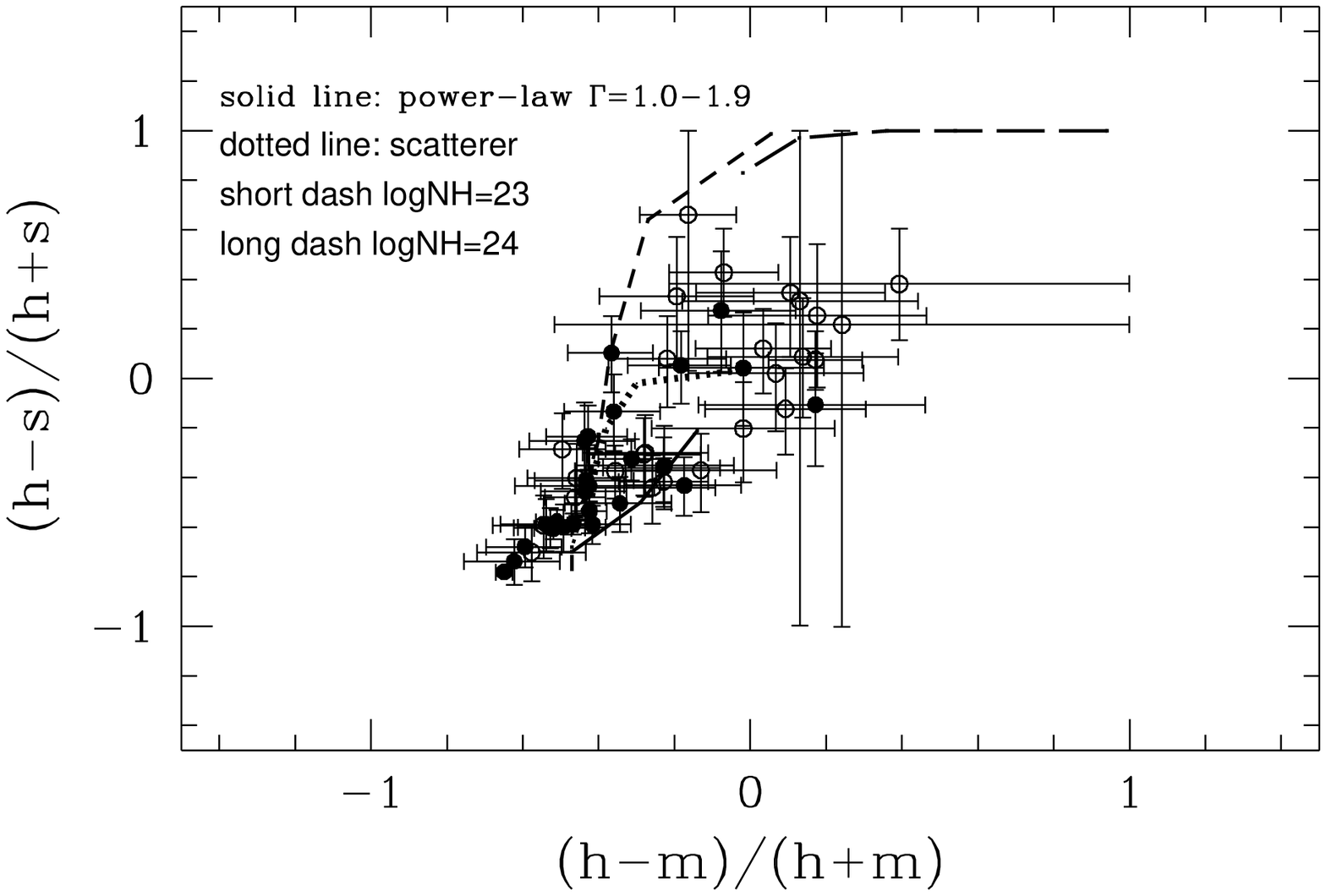}{10truecm}{0}{75}{75}{-250}{-120}
\caption{The X-ray colour-colour diagram.
 The s, m and h bands correspond to 0.7-2, 2-5 and 5-10 
 keV respectively. Open and filed circles denote
 sources with detections in ASCA  and ROSAT respectively.
 Sources with no detection in the 0.7-2 and 2-5 keV bands have been omitted 
 for simplicity.}
\end{figure}

\vfil

 A key question is whether our sources present 
 hard spectra due to large amount 
 of intrinsic absorption or whether they 
 are intrinsically hard. 
 For example, bremsstrahlung emission from 
 an ADAF would produce a spectrum with $\Gamma=1.4$ 
 in our band (Narayan 1999). 
 A pure reflection spectrum eg in 
 the case of a Compton-thick Seyfert galaxy 
 could produce an even flatter spectrum. 
 In order to investigate, in more detail, the 
 spetral properties of our sources and 
 in particular the role of absorption,
 we plot the X-ray colour-colour diagram (Fig. 3).
 Here we compare the softness ratio (h-s/h+s) 
 versus the hardness ratio (h-m/h+m) 
 where s,m and h are the counts 
 in the 0.7-2, 2-5 and 5-10 keV band respectively. 
 A wide range of spectral properties is evident 
 on Fig. 3. Absorbed model spectra  
  consisting of a power-law of $\Gamma=1.9$ (the typical 
 AGN spectrum) and a column  
 $N_H=10^{24}$ and $N_H=10^{23}$ $\rm cm^{-2}$ 
  are denoted with a long and a short dash respectively;
 the above model spectra are evolved from redshift 
 z=0 to redshift z=2  progressively getting softer as the 
 K-correction moves the absorption to softer energies 
 outside the ASCA band. 
 The solid line represents a power-law spectrum 
 with no absorption; the softest end of the line corresponds 
 to $\Gamma=1.9$ while the  hardest point to $\Gamma=1.0$.
 Finally, the dotted line corresponds to a 
 scattering model. Here, we assume that the hard X-ray emission 
 is covered by an obscuring screen of $N_H=10^{23}$.
 10\% of the X-ray emission is scattered along the line 
 of sight due to an electron scattering medium. 
 This model is typical of intermediate Seyfert galaxies 
 (eg Seyfert 1.8 -1.9) in the local Universe
 (eg Turner et al. 1997). Again we evolve our model from redshift 
 z=0 to z=2. We see that the scatterer model provides 
an excellent description for a large number of our sources. 
 Della Ceca et al. (1999) have reached similar conclusions 
 studying the spectral properties of 
 a sample of ASCA sources  detected in the softer 2-10 keV band.

\section{Conclusions}
We report the first results from the hard (5-10 keV) 
 ASCA GIS survey SHEEP. The purpose of our survey is to 
 detect bright, nearby examples of the hard populations 
 which constribute the largest farction 
 of the XRB. We have detected 69 sources in 39 $\rm deg^2$.
 32 of these sources have ROSAT counterparts and therefore 
 have a rather secure optical counterpart.  
  Already existing literature classifications show 
 that 12 of the ROSAT sources are associated 
 with type-1 AGN (ie Seyfert-1 and QSOs). 
 Some of the above  sources present very hard X-ray colours 
 in the 2-10 keV band, comparable or even harder than  
 the spectrum of the XRB. 
 Such obscured QSOs at high redshift have been also found in the 
 BeppoSAX HELLAS survey. 
 Furthermore, the X-ray colour-colour analysis shows that 
 the majority of our sources can be described with 
 obscured AGN models with large amounts of absorption 
 ($>10^{22}\rm cm^{-2}$) but where  small amounts of soft X-ray emission 
 are also present. The soft X-rays are possibly 
 arising from scattering of the primary 
 X-ray continuum along the line of sight on a warm medium.
 This ``scatterer'' model is in line 
 with the Seyfert unification scheme scenaria.

\end{document}